\documentclass[showpacs,prl]{revtex4}
\usepackage{graphicx}
\begin{document}
\title{Charge order in photosensitive Bi$_{0.4}$Ca$_{0.6}$MnO$_3$ films\\}

\author{C.S. Nelson}
\affiliation{National Synchrotron Light Source, Brookhaven National Laboratory, Upton, NY  11973-5000}
\author{R.M. Kolagani}
\thanks{formerly M. Rajeswari}
\affiliation{Department of Physics, Astronomy, and Geosciences, Towson University, Towson, MD  21252}
\author{M. Overby}
\affiliation{Department of Physics, Astronomy, and Geosciences, Towson University, Towson, MD  21252}
\author{V.N. Smolyaninova}
\affiliation{Department of Physics, Astronomy, and Geosciences, Towson University, Towson, MD  21252}
\author{R. Kennedy}
\affiliation{Department of Physics, Astronomy, and Geosciences, Towson University, Towson, MD  21252}

\date{\today}
\begin{abstract}
We report the structural and electronic characterization of the charge order phase in Bi$_{0.4}$Ca$_{0.6}$MnO$_3$ films, in which photoinduced resistivity changes have been observed at temperatures approaching room temperature.  In all films, lattice distortions associated with the charge order are observed, and both the wavevectors and displacements of the distortions are in the plane of the film.  Films under compressive and tensile strain are observed to have different resonant x-ray scattering characteristics--- a result that may shed light on the mechanism responsible for the photosensitivity exhibited by this material.
\end{abstract}
\maketitle

Perovskite manganites of formula R$_{1-x}$M$_{x}$MnO$_3$ (where R and M are trivalent rare-earth and divalent alkaline cations, respectively) are of interest for the wide variety of properties that they exhibit, including their sensitivity to perturbations such as magnetic field (e.g., the colossal magnetoresistance effect\cite{jin}) and photon irradiation (e.g., photosensitivity), with photon wavelengths ranging from the near infrared to the x-ray.\cite{miyano,valery}  Such fascinating behaviors are believed to arise from the interplay of lattice, charge, orbital, and magnetic degrees of freedom, and therefore present a great challenge to those in pursuit of a thorough understanding of these complicated materials.

Bi$_{1-x}$Ca$_x$MnO$_3$ is an example of a perovskite manganite that has been the focus of recent work due to its photosensitivity.\cite{smolbulk,smolfilm}  Remarkably, photosensitivity--- in the form of a reduced resistivity upon exposure to visible light--- has been observed at high temperatures (up to T$\sim$250 K) in Bi$_{0.4}$Ca$_{0.6}$MnO$_3$ films.\cite{smolfilm}  The combination of the effect being exhibited by a film, and the high temperatures at which it is observed, immediately suggests the potential of this material for use in photonic devices.  However, many questions about the photosensitivity remain.  For example, the mechanism responsible for the photosensitivity is unknown, both in the Bi$_{0.4}$Ca$_{0.6}$MnO$_3$ films as well as other manganites that exhibit photosensitivity.  Proposed mechanisms focus on the effects of the photon irradiation on the charge order (CO), since the CO phase is the ground state phase of the material when it is photosensitive, and there is speculation that local melting of the CO leads to the photosensitivity.\cite{smolbulk,satoh}  Whether this melting arises from a structural (e.g., lattice relaxation) or electronic (e.g., loss of Mn charge disproportionation) mechanism is a subject of debate.  Another question concerns the effects of thin film properties such as thickness and strain on the photosensitivity, since they have been reported to correlate with the magnitude of the photoinduced resistivity changes in Bi$_{0.4}$Ca$_{0.6}$MnO$_3$ films.\cite{smolfilm}

In order to address these important questions, we report a study of the structural and electronic properties of the CO phase in Bi$_{0.4}$Ca$_{0.6}$MnO$_3$ films.  We find that the wavevectors of the lattice distortions associated with the CO are in the plane of the film for all films investigated.  The lattice distortions also have displacements in the plane of the film.  The displacements are strictly transverse under tensile strain, but under compressive strain, have an additional, longitudinal component.  We also observe, using resonant x-ray scattering techniques, different electronic characteristics of the CO phase in films under tensile versus compressive strain.  Specifically, a film under tensile strain exhibits a resonant enhancement consistent with the presence of Mn charge disproportionation or charge anisotropy, while a film under compressive strain does not.  Altogether, our results underscore the importance of the lattice degree of freedom in Bi$_{0.4}$Ca$_{0.6}$MnO$_3$ films, and further, suggest that the structural properties of the CO phase play the dominant role in the observed photosensitivity.

The films were grown using pulsed laser deposition with two different substrates:  SrTiO$_3$(100) (STO) and LaAlO$_3$(100) (LAO).  As a representative example of the photosensitivity exhibited by these films, the temperature dependence of the resistivity of a Bi$_{0.4}$Ca$_{0.6}$MnO$_3$ film grown on LAO--- measured with and without visible light illumination--- is shown in Figure 1.  The dramatic effect of the illumination on the resistivity can be clearly seen in this figure.

Comparing the in-plane lattice constants of the two substrates to that of bulk Bi$_{0.4}$Ca$_{0.6}$MnO$_3$ \cite{interpolate}, STO and LAO provide $\sim$1.5--2\% tensile and compressive strain, respectively.  The thicknesses of the films were 60 nm for films grown on each substrate, and 30 nm for a second film grown on a STO substrate.  The x-ray scattering experiments were carried out on beamlines X22A and X22C at the National Synchrotron Light Source at Brookhaven National Laboratory.  Both beamlines are configured in a vertical scattering geometry.  Beamline X22A provides a fixed incident photon energy, which was 10.262 keV, using a single-bounce from a Si(111) monochromator.  Beamline X22C employs a Si(111) double-crystal monochromator, and is tunable, which enabled resonant x-ray scattering measurements to be carried out near the Mn K absorption edge.  A graphite analyzer was used on beamline X22C in order to minimize detection of fluorescence photons from the films.

The structural properties of the films were studied at room temperature.  All of the films were found to have an (002) surface normal orientation, using $\it{Pbnm}$ orthorhombic notation.  Note that while the bulk material is believed to be triclinic at room temperature,\cite{woo} the deviation from an orthorhombic structure is small.  Further, since complete structural refinements of the films are beyond the scope of this paper, orthorhombic notation will be used in what follows.

In order to focus on the CO, which is the primary subject of this paper, a brief summary of previous studies of the CO phase in bulk, electron-doped Bi$_{1-x}$Ca$_{x}$MnO$_3$ is in order.  In work carried out 40 years ago by Bokov {\it et al.},\cite{bokov} a structural transition with concomitant increase in resistivity, for x $<$ 0.85, was believed to coincide with charge ordering on the Mn sites.  Subsequently, both electron\cite{bao,murakami} and x-ray\cite{su} scattering measurements have indicated the presence of a lattice distortion associated with the CO phase.  The lattice distortion is evidenced by the appearance of superlattice peaks with wavevector ($\tau$00), in orthorhombic notation.  The magnitude of $\tau$ depends upon the doping level, x, and $\tau$ values of 0.30, 0.24, 0.25, and 0.22, have been reported for x values of 0.74, 0.76, 0.8, and 0.82, respectively.\cite{bao,murakami,su}  In addition to these scattering studies of the CO phase in Bi$_{1-x}$Ca$_x$MnO$_3$, STM has recently been used to image charge disproportionation in an x = 0.76 sample.\cite{renner}  The charge disproportionation observed in these images is inconsistent with the scattering measurements in that it indicates a Mn$^{3+}$ to Mn$^{4+}$ ratio of 1:1, but the surface sensitivity of STM and/or presence of phase separation were suggested as possible explanations for this discrepancy.  Note that in this paper, CO is used to refer to the high resistivity phase that exhibits a lattice distortion.  Actual charge disproportionation in this phase--- such as that observed using STM\cite{renner}--- will also be addressed, through the electronic characterization of our films.

For our x-ray scattering studies of CO in the films, the samples were mounted inside a displex closed-cycle cryostat, which is equipped with x-ray transparent beryllium domes that are used as a vacuum shroud and a radiation shield, and cooled to a temperature of 10--15 K.  Cooling cycles were carried out at different times with and without x-ray exposure, and no evidence of photosensitivity at x-ray wavelengths was observed.  Specifically, reciprocal space scans along [H42] resulted in the observation of CO superlattice peaks in all of the films--- regardless of x-ray exposure during cooling.  As an example, a scan is shown for the film on LAO, under compressive strain, in Figure 2.  In addition to the three Bragg peaks, CO superlattice peaks are clearly observed on either side of each Bragg peak.  Some second harmonics of the CO superlattice peaks are also observed, and hints of a fourth harmonic can be seen as shoulders on the sides of the Bragg peaks.  We note that some second harmonics could be observed in the thicker film on STO, under tensile strain, but not in the thinner film grown on this substrate.  And in neither of these latter films were any higher harmonics observed.

To characterize the CO phase further, superlattice peaks were searched for in different places in reciprocal space for the two thick films.  Around (242), they were only observed as pairs.  That is, they were seen at (2$\pm\tau$,4,2), but not at (2,4$\pm\tau$,2).  Superlattice peaks were also absent around (004), but superlattice peaks with very weak (a factor of $\sim$100 weaker than the peaks near (242)), and nearly identical, intensities were observed at ($\pm1\mp\tau$,0,4) and (0,$\pm1\mp\tau$,4) for the film grown on LAO.  In contrast, these peaks were absent in the film grown on STO, except when the incident photon energy was at the Mn K edge resonance, which is discussed in more detail below.  The fact that superlattice peaks were observed at ($\pm1\mp\tau$,0,4) and (0,$\pm1\mp\tau$,4) in both films--- whether on- or off-resonance--- indicates the presence of a--b twinning and/or that there are two CO wavevectors:  ($\tau$00) and (0$\tau$0).  In conjunction with the much stronger (2$\pm\tau$,4,2) peaks and the absence of (2,4$\pm\tau$,2) peaks--- again, in both films--- it also suggests that the lattice distortion is primarily transverse, with displacements in the plane of the film.

The temperature dependences of the CO superlattice peaks were studied in all of the films, and the results are displayed in Figure 3.  In this figure, the fitted peak intensities are for superlattice peaks along [H42], they are normalized to a nearby Bragg peak, and the ratio is set to 1 for each film at a temperature of 10 K.  In the inset, the temperature dependences of the ratios of both the CO superlattice peak and its second harmonic for the film grown on LAO, under compressive strain, are displayed, and the close correspondence indicates that the shape of the lattice distortion associated with CO is not changing with temperature in this film.

The CO order parameters show broad transitions in all of the films with the CO superlattice peak intensity decreasing monotonically with temperature above $\sim$50 K.  The film with the highest ordering temperature is the thicker film grown on STO, under tensile strain, which exhibits CO superlattice peaks with nonzero scattering intensity above room temperature.  Comparing the order parameters for the two films on STO, a reduction in thickness is observed to decrease the ordering temperature, but not dramatically.

The lower bounds of the ordering transitions, as determined by the onset of broadening of the CO superlattice peaks, are listed in Table I for all films.  Comparing the value for the film grown on LAO to the photosensitivity data presented in Figure 1, the ordering temperature is seen to be very close to the temperature at which the photosensitivity disappears--- thus tying the structural properties of the CO phase to the photosensitivity.  The values of the CO wavevectors, $\tau$, for the two thick films measured at a temperature of 15 K are also shown in Table I.  We note the trend with the lattice mismatch in that tensile (compressive) strain results in a larger (smaller) wavevector and therefore a shorter (longer) real-space lattice distortion.  In addition, the magnitude of $\tau$ was observed to decrease at the CO transition, which indicates an evolution toward a longer-period, real-space modulation as the CO melts.  It is important also to note that the $\tau$ values listed in Table I were obtained using scans through (1-$\tau$,0,4) superlattice peaks.  Since $\gamma$, which is the angle between the in-plane a and b lattice vectors, is expected to deviate the most from 90$^{\circ}$,\cite{bokov} reciprocal space positions with H or K equal to zero are least affected by the deviation from an orthorhombic structure.  By comparing the $\tau$ values obtained from scans through (1-$\tau$,0,4) and ($\tau$,4,2), we can determine $\gamma$ values, which are listed in the table.  Note that the type of strain is clearly manifested in the $\gamma$ values, in that compressive (tensile) strain results in $\gamma < (>) 90^{\circ}$.  $\tau$ and $\gamma$ values were impossible to obtain for the thinner film grown on STO due to the weak scattering intensity at (1-$\tau$,0,4).

Another focus of our studies was the electronic characterization of the CO phase, which we carried out by using resonant x-ray scattering techniques.  As has been demonstrated in a number of publications over the past several years,\cite{nakamura,zim,staub,grenier,wilkins,ohwada} charge disproportionation in transition metal oxide compounds gives rise to a K edge resonant enhancement at the CO wavevector.  Since this scattering can be quite weak and therefore dominated by scattering due to the lattice distortion associated with the CO, we studied the scattering at a position in reciprocal space in which the scattering contribution from the latter was found to be negligible:  near the (004) Bragg peaks.

Figure 4 displays the energy dependence of the scattering at the CO wavevector and in the nearby background for the film grown on STO, under tensile strain.  An enhancement in the scattering intensity above the fluorescence that leaks through the analyzer can be clearly observed at the CO wavevector, near an incident photon energy of 6.555 keV.  The inset, which shows the difference between the two measurements, underscores this point, and a resonant enhancement of $\sim$4 s$^{-1}$ can be seen.  This enhancement was not observed for the film grown on LAO, under compressive strain.

Next, in Figure 5, reciprocal space scans through the CO superlattice peak measured at different incident photon energies for films on STO and LAO are displayed.  Focusing first on the film under tensile strain (Figure 5a), the scattering at the CO wavevector is shown to be purely resonant in nature, in that the peak is absent when the incident photon energy is not at the resonant energy.  This observation leads to two conclusions:  charge disproportionation and/or charge anisotropy occurs in this film,\cite{origin} and the lattice distortion is transverse with in-plane distortions--- as mentioned earlier.  More explicitly, the latter result follows from the leading order ${\bf Q}\cdot\vec{\delta}$ amplitude for scattering from a lattice distortion with displacement $\vec{\delta}sin(\vec{\tau}\cdot{\bf r})$, which is zero for ${\bf Q}$=(0,1-$\tau$,4) and $\vec{\delta}\parallel$ \^{a}.  In Figure 5b, which shows similar reciprocal space scans for the film grown under compressive strain, the opposite conclusion is reached:  there is no evidence of charge disproportionation in this film, and the nonresonant scattering intensity indicates the presence of a distortion in addition to the dominant, transverse distortion.  The two possibilities for this second distortion are a transverse distortion with out-of-plane displacements, or a longitudinal distortion.  Since the intensities of these CO superlattice peaks are observed to be proportional to ${\bf Q}\cdot a^*$ or ${\bf Q}\cdot b^*$ and not ${\bf Q}\cdot c^*$, the second distortion appears to be longitudinal.

The different characteristics of the CO phase in the films grown under tensile versus compressive strain are striking.  The ordering temperatures, wavevectors, and directions of in-plane displacements of the lattice distortions associated with the CO are affected by the type of strain.  More surprising, however, is the absence of evidence of charge disproportionation in the film grown under compressive strain.  One explanation for this absence is that the resonant scattering is below the detection limits of the measurements.  It is unclear why this would be so, when compared to the film grown under tensile strain, which has broader CO superlattice peaks and yet clear evidence of a resonant enhancement.  The other explanation--- that charge disproportionation does not occur in the film grown under compressive strain--- suggests that the structural properties of the CO phase play the dominant role in the observed photosensitivity in this material.  We note that films of a given thickness under compressive strain were observed to exhibit larger photoinduced resistivity changes compared to films under tensile strain,\cite{smolfilm} which both underscores this conclusion and raises the possibility that the additional longitudinal distortion in the former films may enhance the photosensitivity.  This possibility will be the subject of future studies.

In conclusion, we have carried out the structural and electronic characterization of the CO phase in Bi$_{0.4}$Ca$_{0.6}$MnO$_3$ films.  We have observed differences in the lattice distortions associated with the CO, in films grown under comparable amounts of compressive versus tensile strain.  An absence of charge disproportionation in a film grown under compressive strain was also observed, which is consistent with the structural properties of the CO phase playing the key role in its photosensitivity.

CSN thanks Valery Kiryukhin for helpful discussions.  Work carried out at the National Synchrotron Light Source, Brookhaven National Laboratory, was supported by the U.S. Department of Energy, Division of Materials Sciences and Division of Chemical Sciences, under Contract No. DE-AC02-98CH10886.  Work at Towson University was supported by the NSF under grants DMR-0348939 and DMR-0116619.

\pagebreak

\begin{table}
\caption{Bi$_{0.4}$Ca$_{0.6}$MnO$_3$ films}
\begin{ruledtabular}
\begin{tabular}{cccccc}
& thickness (nm) & T$_{co}$ (K) & $\tau$ (rlu)\footnotemark[1] & $\gamma$ ($^{\circ}$) \\  \hline
on STO & 30 & 300 & & \\
on STO & 60 & 325 & 0.29 $\pm$ 0.01 & 90.15 $\pm$ 0.1 \\
on LAO & 60 & 250 & 0.27 $\pm$ 0.01 & 89.3 $\pm$ 0.1 \\
\end{tabular}
\end{ruledtabular}
\footnotetext[1]{values obtained at T = 15 K, using scans through (1-$\tau$,0,4) CO superlattice peaks}
\end{table}
 
\begin{figure}
\includegraphics[width=7.5cm]{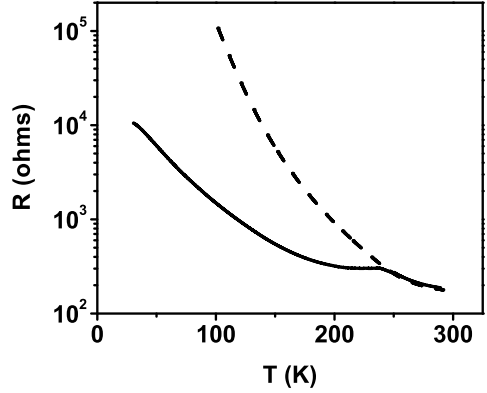}
\caption{Temperature dependence of the resistivity of a Bi$_{0.4}$Ca$_{0.6}$MnO$_3$ film grown on LAO.  The solid line represents R(T) under illumination with visible light ($\sim$500 nm wavelength), and the dashed line represents R(T) without illumination.  Both data sets were taken upon sample warming.}
\end{figure}

\begin{figure}
\includegraphics[width=6cm]{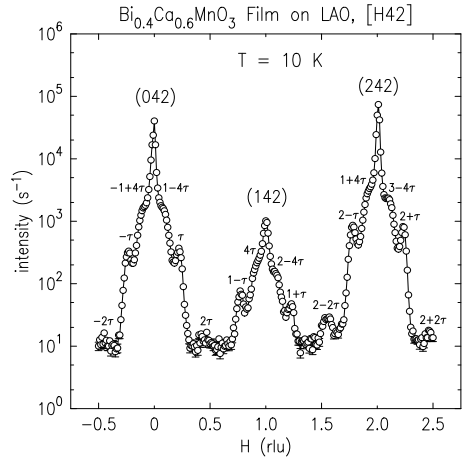}
\caption{Reciprocal space scan through CO superlattice peaks.}
\end{figure}

\begin{figure}
\includegraphics[width=6cm]{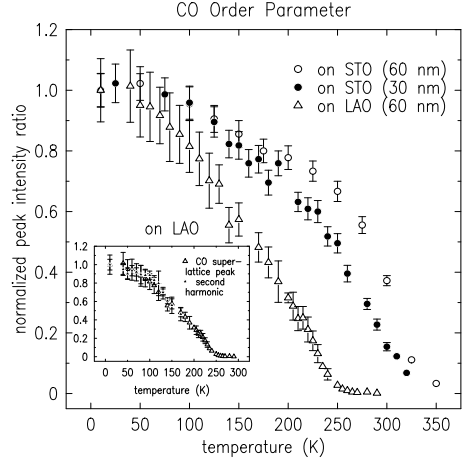}
\caption{CO order parameter for films on STO and LAO substrates.  Inset displays the temperature dependences of both the CO superlattice peak and its second harmonic for the film on LAO.}
\end{figure}

\begin{figure}
\includegraphics[width=6cm]{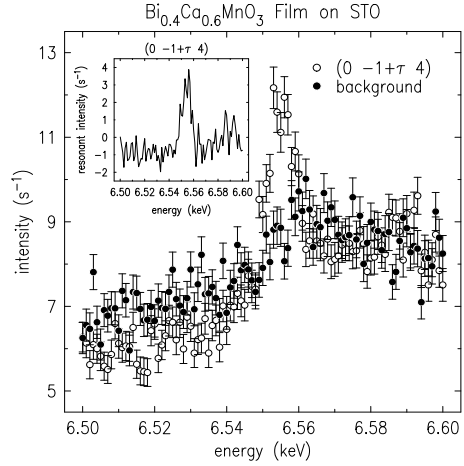}
\caption{Energy dependence of the scattering intensity at the CO wavevector ($\circ$) and in the background ($\bullet$).  The inset shows the difference between the two sets of measurements, and errorbars have been excluded for clarity.}
\end{figure}

\begin{figure}
\includegraphics[width=6cm]{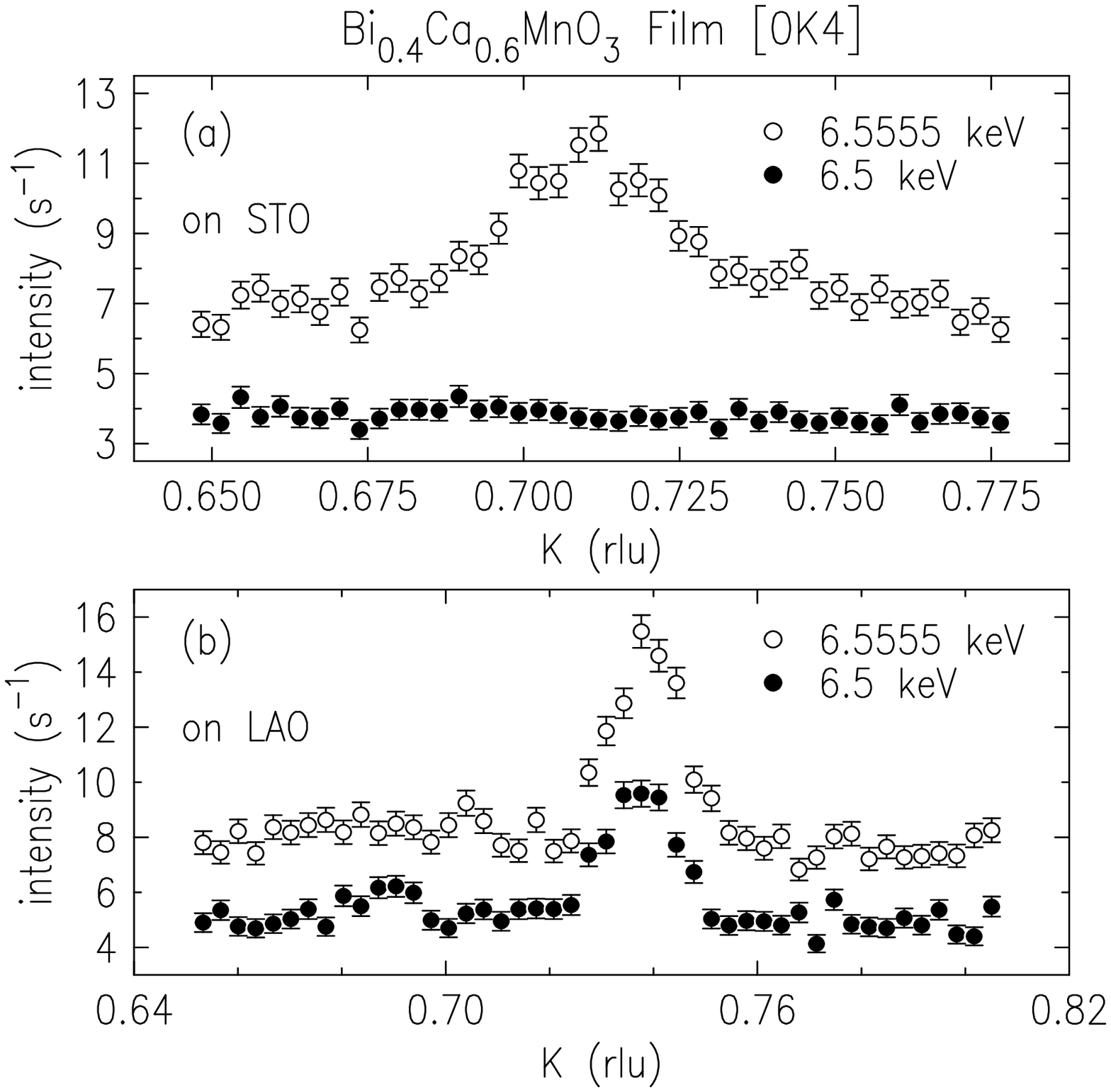}
\caption{Reciprocal space scans, on- ($\circ$) and off- ($\bullet$) resonance, through the (0,1-$\tau$,4) superlattice peaks, for films on STO (a) and LAO (b).}
\end{figure}


\begin{thebibliography}{99}
\bibitem{origin} Both charge disproportionation and charge anisotropy give rise to a resonant enhancement.  Measurements of the polarization and azimuthal dependences of the scattering can be used to separate the two contributions, but such measurements are impossible to carry out with the extremely weak resonant scattering intensity and the given orientation of the films.
\end{thebibliography}
\end{document}